\documentclass[a4paper,fleqn,usenatbib,letters]{mnras}

\usepackage[T1]{fontenc}
\usepackage{ae,aecompl}

\usepackage{graphicx}	
\usepackage{amsmath}	
\usepackage{amssymb}	

\title[Estimating stellar wind parameters]{Estimating stellar wind parameters from low-resolution magnetograms}

\author[M.Jardine et al.]{M. Jardine$^1$ \thanks{E-mail: mmj@st-andrews.ac.uk}, A. A. Vidotto$^2$ and V. See$^1$\\
$^1$ SUPA, School of Physics \& Astronomy, University of St Andrews, North Haugh, St Andrews,  KY16 9SS, UK\\
$^2$ School of Physics,  Trinity College Dublin, The University of Dublin, Dublin-2, Ireland\\
}

\date{Accepted XXX. Received YYY; in original form ZZZ}

\pubyear{2016}

\begin{document}
\label{firstpage}
\pagerange{\pageref{firstpage}--\pageref{lastpage}}
\maketitle

\begin{abstract}
Stellar winds govern the angular momentum evolution of solar-like stars throughout their main-sequence lifetime. The efficiency of this process depends on the geometry of the star's magnetic field. There has been a rapid increase recently in the number of stars for which this geometry can be determined through spectropolarimetry. We present a computationally efficient method to determine the 3D geometry of the stellar wind and to estimate the mass loss rate and angular momentum loss rate based on these observations. Using solar magnetograms as examples, we quantify the extent to which the values obtained are affected by the limited spatial resolution of stellar observations. We find that for a typical stellar surface resolution of 20$^{\rm o}$-30$^{\rm o}$, predicted wind speeds are within 5$\%$ of the value at full resolution. Mass loss rates and angular momentum loss rates are within 5-20$\%$. In contrast, the predicted X-ray emission measures can be under-estimated by 1-2 orders of magnitude, and their rotational modulations by 10-20$\%$.
\end{abstract}





\section{Introduction}

The angular momentum evolution of solar-like stars is governed by the action of their winds and in particular by the interaction between the hot, escaping gas and the stellar magnetic field \citep{parker58}.  Studies of these stellar winds are hampered however by the low density of the wind plasma which makes direct detection difficult \citep{wood_asterospheres_05}. Often the mass loss can only be inferred by studying the rotational distributions of samples of coeval stars in young clusters \citep{irwin2011,delorme2011}. The strength of the stellar magnetic field, which principally determines the extent of the level arm that the wind may apply, is clearly an important parameter in determine the instantaneous torque applied by the wind \citep{weber67}. The field topology is also important however as only open field lines can support a wind \citep{mestel68,mestel87}. 

The {\em open} flux of magnetic field depends crucially on the geometry of the magnetic field. Over the last decade, advances in spectropolarimetry have provided surface magnetograms for a wide range of stellar masses and ages through the technique of Zeeman-Doppler imaging \citep{donati_review_2009}. These underpin theoretical efforts to model the structure and evolution of the coronae and winds of these stars \citep{ vidotto2009,Cranmer_Saar_2011,matt2012,vidotto2013,cohen_grid2014,vidotto2014,vidotto_trends2014,matt2015,reville_rss2015,reville2015,vidotto2015,see_confusagram_2016}. Advances in the application of 3D MHD wind models to this data have allowed us to study the unusually powerful winds of low mass stars \citep{vidotto_v374peg_10}, young active stars \citep{cohen_abdor_10}, the role of non-potential field \citep{jardine2013}, the impact of stellar winds on exoplanetary magnetospheres \citep{cohen_HD189733_2011,vidotto_tauBoo2012} and the relationship between mass loss rates and X-ray fluxes \citep{vidotto_wood2016}.


\begin{figure}
\begin{center}
	\includegraphics[width=40mm]{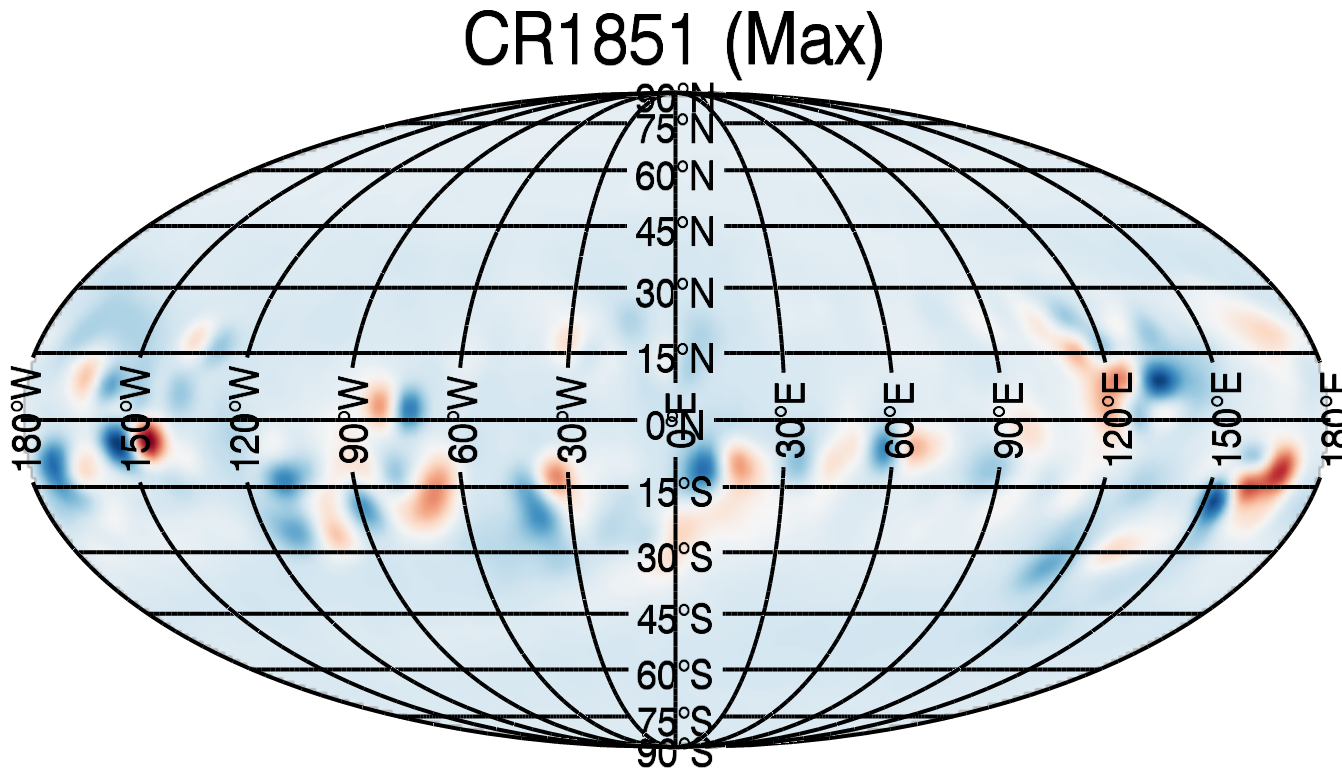}
	\includegraphics[width=40mm]{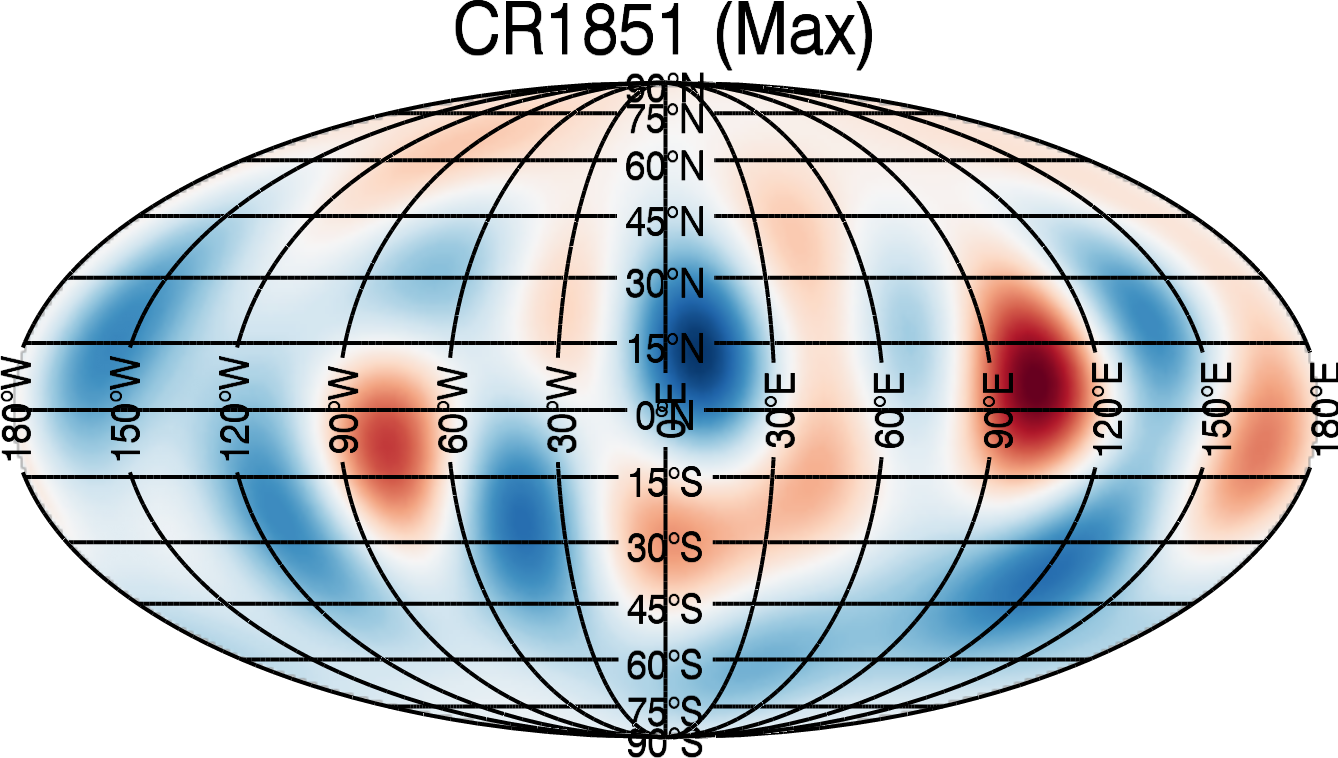}\\
	\includegraphics[width=41mm]{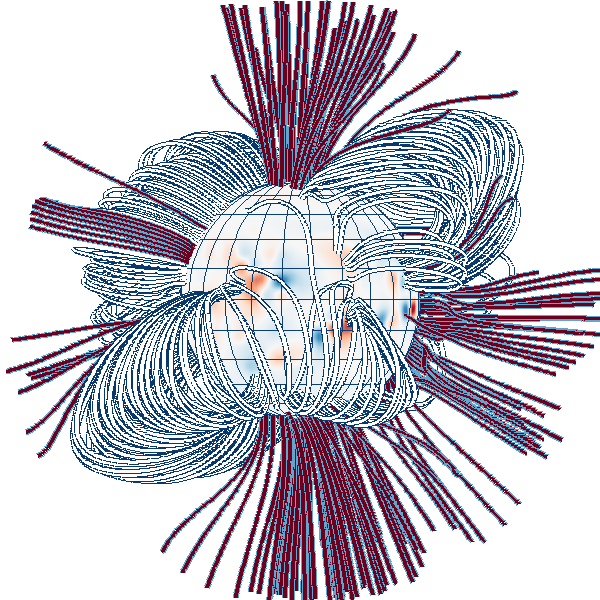}
	\includegraphics[width=41mm]{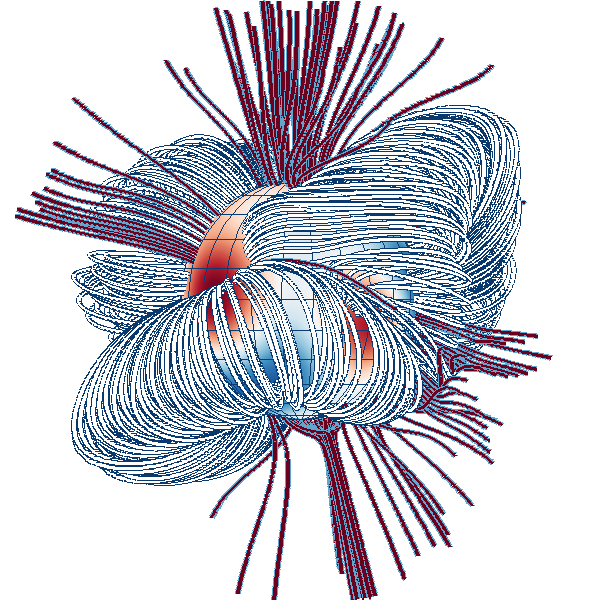}
\caption{Top: Surface magnetograms for Carrington rotation 1851 (close to cycle maximum). The map is reconstructed for a maximum spherical harmonic degree of (left) $\ell_{\rm max} = 63$ and (right) $\ell_{\rm max} = 5$ corresponding to surface spatial scales of $~3^{\rm o} $ and $30^{\rm o}$ respectively. Colourbars are set to $\pm 200$G (left) and $\pm 30$G (right) . Bottom: The corresponding field extrapolations, with wind-bearing (open) field lines coloured red. The overall structure of the largest fieldlines is very similar. }
\label{surf_maps}
\end{center}
\end{figure}


Zeeman-Doppler imaging has some limitations, however. It is relatively insensitive to flux in dark (spotted) regions. If this missing flux is a large contribution to the total stellar magnetic flux, its neglect may have a significant effect on the predicted X-ray emission measure \citep{arzoumanian_10,johnstone_spots_10}. The effective surface resolution is also limited and while at best may be of order 5$^{\rm o }$ it is typically 20$^{\rm o }$-30$^{\rm o }$. Since the polarisation signature of small-scale structures may cancel out, as much as 85$\%$-95 $\%$ of the surface flux may be missed \citep{reiners2009}. A consistent picture is emerging, however, of the effect of this missing flux.  By adapting solar magnetograms, \citet{garraffo2013} re-distributed large-scale flux onto smaller lengthscales, thus reducing the open flux, while \citet{lang2014} artificially added a carpet of small-scale field to Zeeman-Doppler maps of 12 M dwarfs. In the first case, the open flux was reduced, and hence the wind properties varied. In the second case, however, the open flux was unaffected. While these studies have shown the robustness of the stellar wind to the presence of unresolved flux, they demonstrate the much more sensitive response of the predicted X-ray emission.  


\begin{figure}
\begin{center}
	\includegraphics[width=75mm]{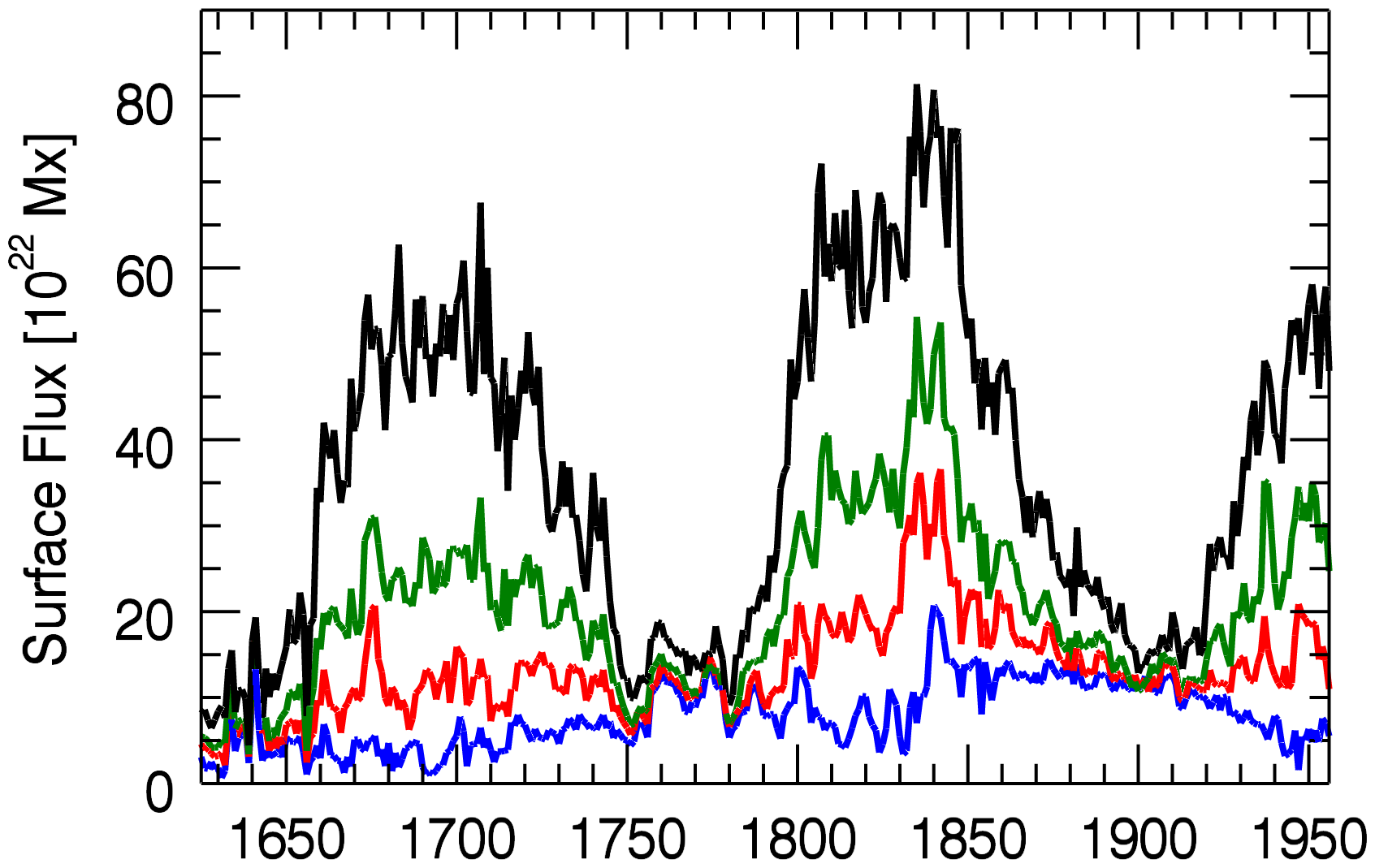}
	\includegraphics[width=75mm]{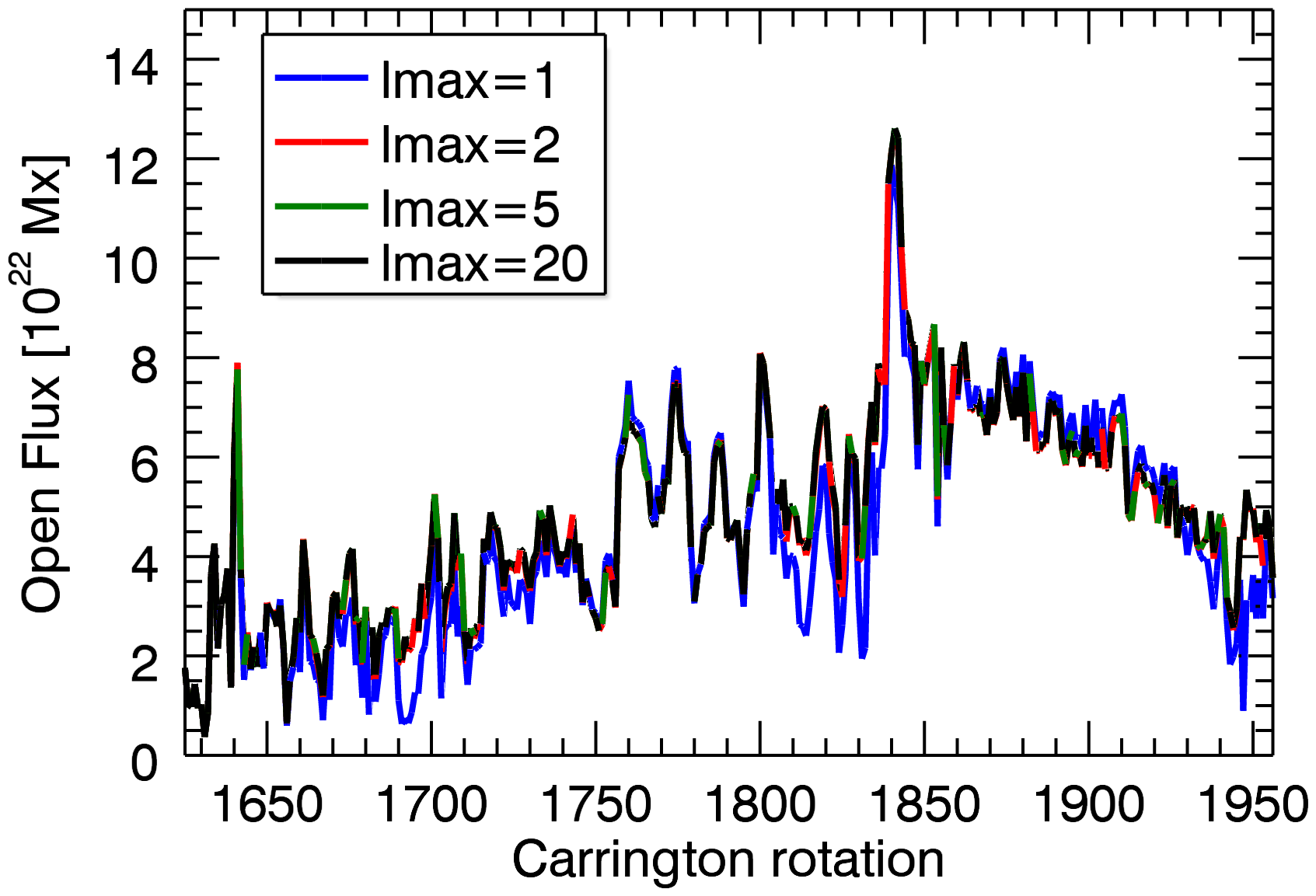}
\caption{Magnetic flux as a function of time for surface magnetograms that have been truncated at some maximum value $\ell_{\rm max}$ in the spherical harmonic expansion (corresponding to minimum surface spatial scales of $180^{\rm o}/\ell_{\rm max}$). The top panel shows the surface flux $\Phi_{\rm surf} = \oint_{r_\odot} |B_r(r_\odot)|dS$, and the bottom panel shows the open flux $\Phi_{\rm open} = \oint_{r_s} |B_r(r_s)|dS$ (i.e. the flux at the radius where the field becomes open).}
\label{B_time}
\end{center}
\end{figure}


As the number of stars whose surface fields has been mapped grows, so does the scope of the models of the coronae and winds of these stars. While fully 3D MHD models provide insight into the nature of these winds, it is clear that a more computationally efficient method of assessing the nature of these winds is needed. In the case of the solar wind, the availability of {\it in situ} measurements has made it possible to develop such an empirical wind model that is calibrated to reproduce the velocity of the solar wind at Earth. The WSA model \citep{wang1990,arge2000} requires only the surface magnetogram as an input. The output is a fully 3D wind model, providing the local wind speed for any location within the solar wind. This model maps wind speeds directly to the degree of expansion of individual flux tubes and so is completely determined by the geometry of the magnetic field. This is typically calculated using a {\em Potential Field Source Surface}  method that assumes the field is potential and is opened at some specified radius  \citep{altschuler69}. The location of this opening radius (the source surface) is a free parameter of the model that is calibrated using solar eclipse images. Using high-resolution solar magnetograms \citet{cohen2015} compared this model with the output of a fully 3D MHD treatment and found differences in arrival times at Earth of more than five hours (out of a travel time of order 3 days) for only $20-40\% $ of field lines. He also concluded that doubling the resolution of the magnetograms from 2$^{\rm o }$ to 1$^{\rm o}$ has little effect on the predicted wind speeds.

While the WSA model has been developed for the Sun, and underlies many space weather and solar wind studies \citep{pinto_solarcycle_11,gressl2014,pinto2016}, it has also been used in conjunction with stellar magnetograms obtained from ZDI in order to predict the impact of stellar winds on exoplanets \citep{fares_2010,fares_2012,see2014}. As more exoplanets are discovered around stars with a greater range of masses and ages, there is clear demand for an efficient but reliable method of estimating wind speeds and mass loss rates of a large number of stars. In this paper, we quantify the reliability of the WSA method when used with stellar magnetograms which have a much lower resolution than the solar magnetograms for which the method was developed.


\section{Method}

We take solar magnetograms obtained from the US National Solar Observatory, Kitt Peak, over two solar cycles, from February 1975 (CR1625) to April 2000 (CR1962). Fig. \ref{surf_maps} shows one example, taken from close to solar maximum. Since the surface field can be expressed as a sum of spherical harmonics, it is possible to truncate this sum at any order of the expansion. A magnetogram where a maximum order $\ell_{max}$ has been used therefore corresponds to a minimum spatial scale at the stellar surface of $180^{\rm o}/\ell_{max}$. We extrapolate the field using the Potential Field Source Surface method, with a source surface at 2.5 $r_\odot$ \citep{riley_PFSS_06}. 


\subsection{Modelling the magnetic field}
\label{potential_section}


\begin{figure}
\begin{center}
	\includegraphics[width=70mm]{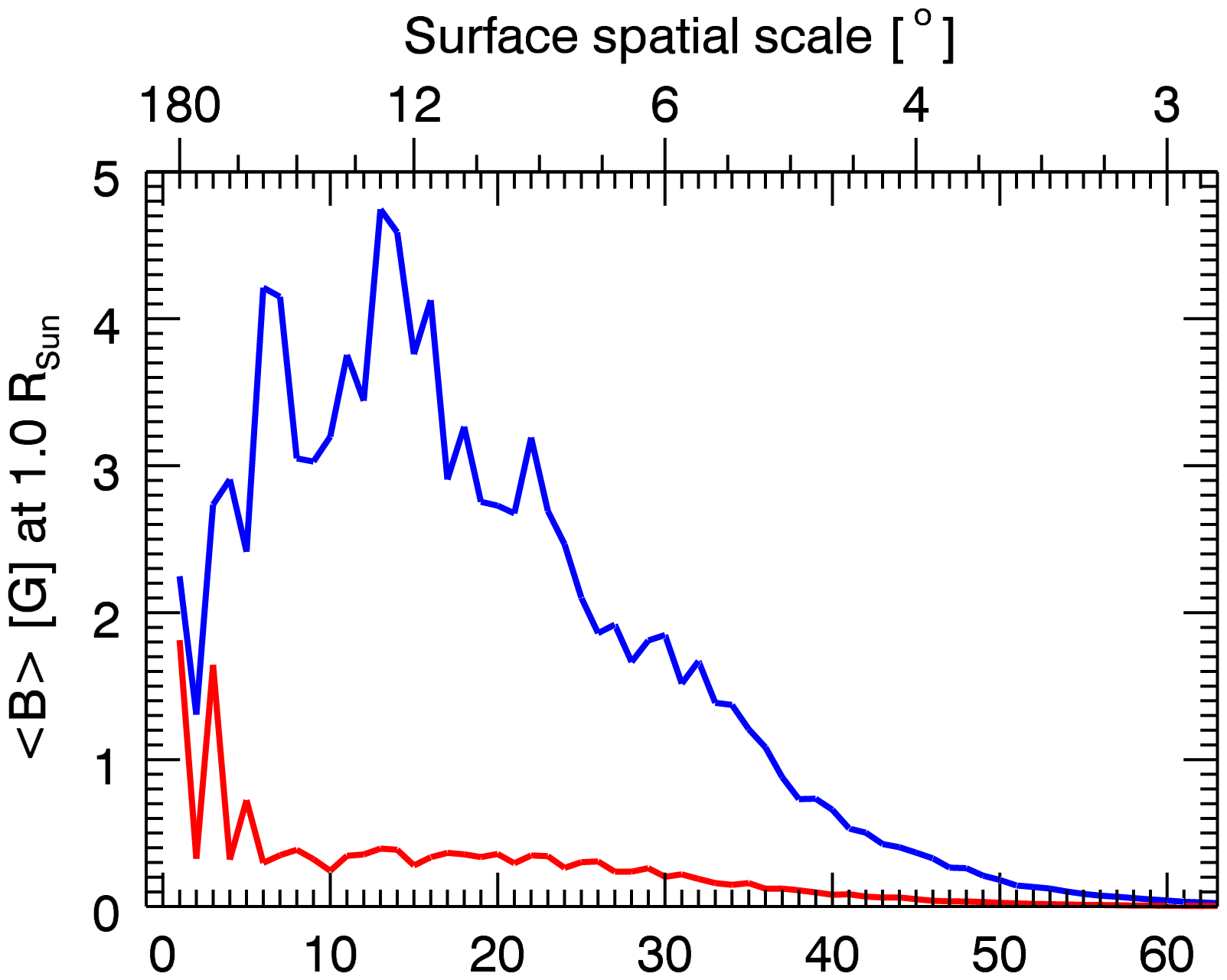}
	\includegraphics[width=74mm]{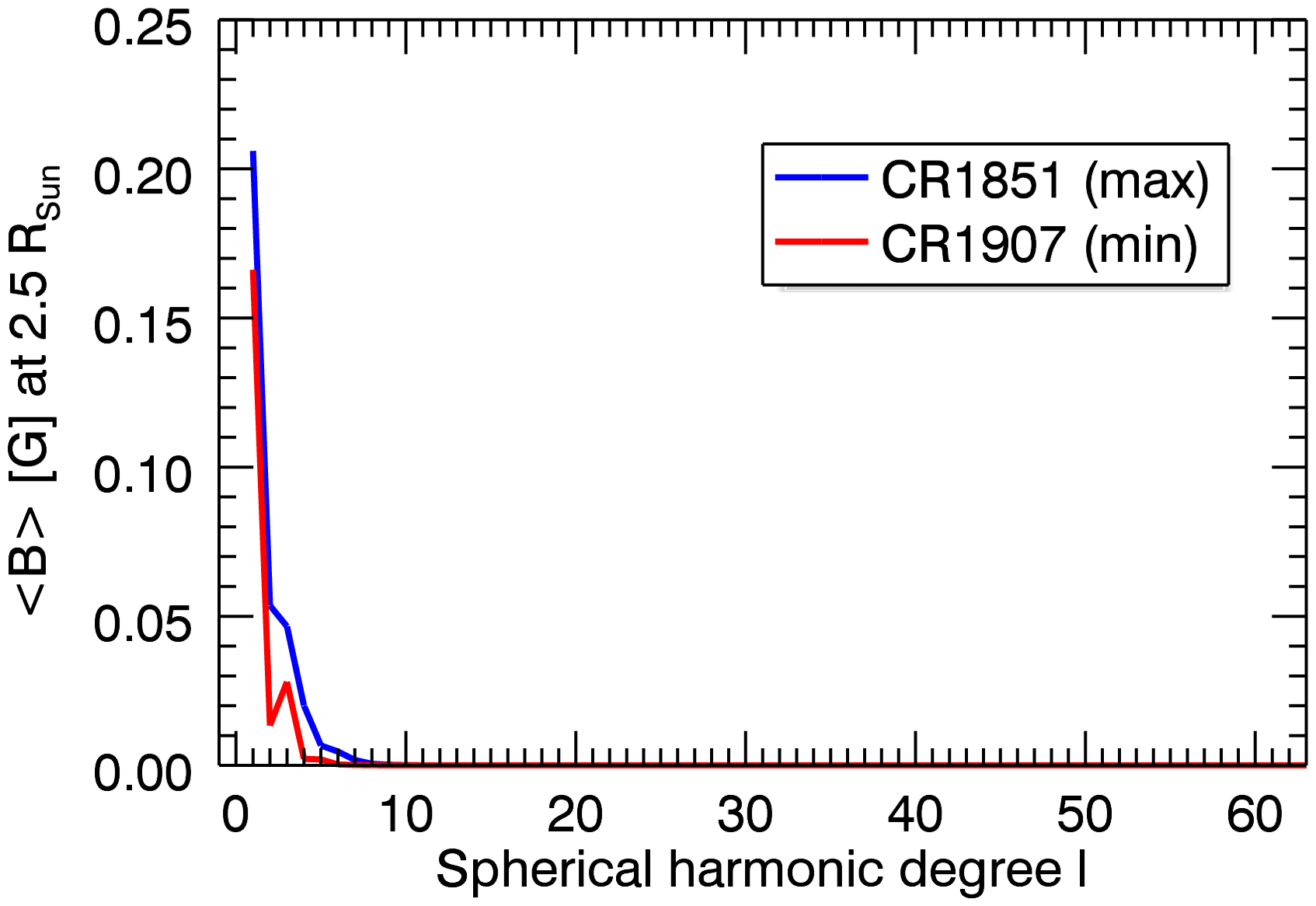}
\caption{Mean flux density  $<B>$ at each individual $\ell$ degree in the spherical harmonic expansion (corresponding to surface spatial scales of $180^{\rm o}/\ell$ ). This is defined as $<B> = \Phi /4\pi r^2$ where the flux through some radius $r$ is $\Phi = \oint_{r} |B_r|dS$. In the top panel, this surface is $r = r_\odot$, whereas in the bottom panel we choose the start of the wind-bearing region $r = r_s$.}
\label{B_ell}
\end{center}
\end{figure}


We assume that the field is potential and divergence-free, such that if $\underline{B} = -\underline{\nabla} \psi$ , then $\psi$ satisfies Laplace's equation $\underline{\nabla}^2 \psi=0$ with solution in spherical co-ordinates $(r,\theta,\phi)$
\begin{equation}
B_r =  \sum^N_{l=1}\sum^l_{m=-l} 
                   [  l a_{lm}r^{l-1} - (l+1) b_{lm} r^{-(l+2)} ] P_{lm}(\cos \theta)e^{im\phi} 
\label{br_pot}
\end{equation}
\begin{equation}
B_\theta   =   -   \sum^N_{l=1}\sum^l_{m=-l} 
                    [  a_{lm}r^{l-1} -  b_{lm} r^{-(l+2)} ] \frac{d}{d\theta}P_{lm}(\cos \theta)e^{im\phi} 
\label{btheta_pot}
\end{equation}
\begin{equation}
B_\phi   =  - \sum^N_{l=1}\sum^l_{m=-l} 
                   [  a_{lm}r^{l-1} -  b_{lm} r^{-(l+2)} ] \frac{P_{lm}(\cos \theta)}{\sin \theta} im e^{im\phi}   
\label{bphi_pot}
\end{equation}
where all radii are scaled to a stellar radius and the associated Legendre polynomials are denoted by $P_{lm}$.  The
two unknowns are the coefficients $a_{lm}$ and $b_{lm}$. One of these can be determined by imposing the
radial field at the surface from the Zeeman-Doppler maps. The second is determined by imposing the condition that at the source surface ($r=r_s$), the field is purely radial, such that $B_\theta (r_s) = B_\phi (r_s) = 0$. We use a code originally developed by \citet{vanballegooijen98} (see also \citet{jardine02structure}). 
%
\begin{figure}
\begin{center}
	\includegraphics[width=70mm]{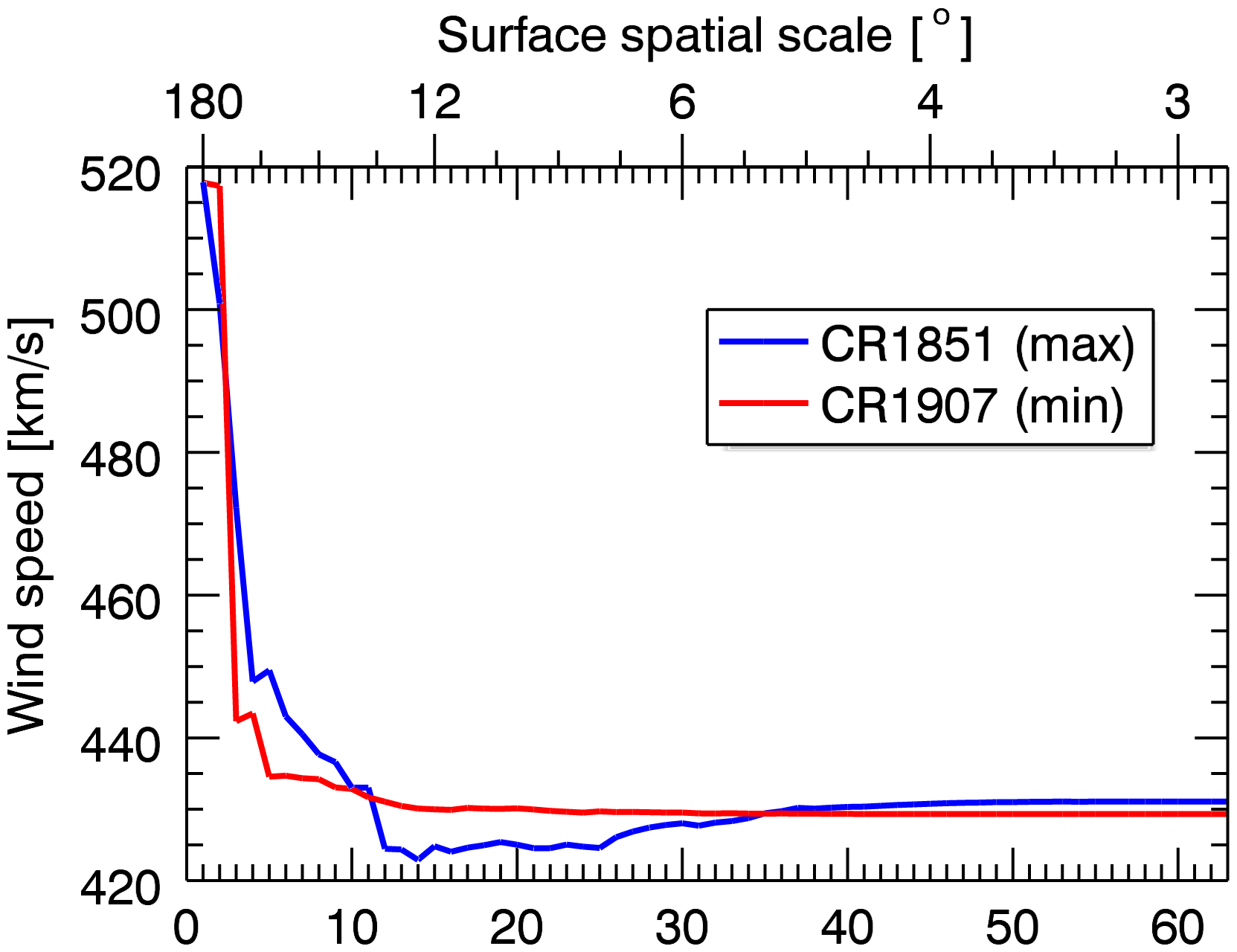}
	\includegraphics[width=70mm]{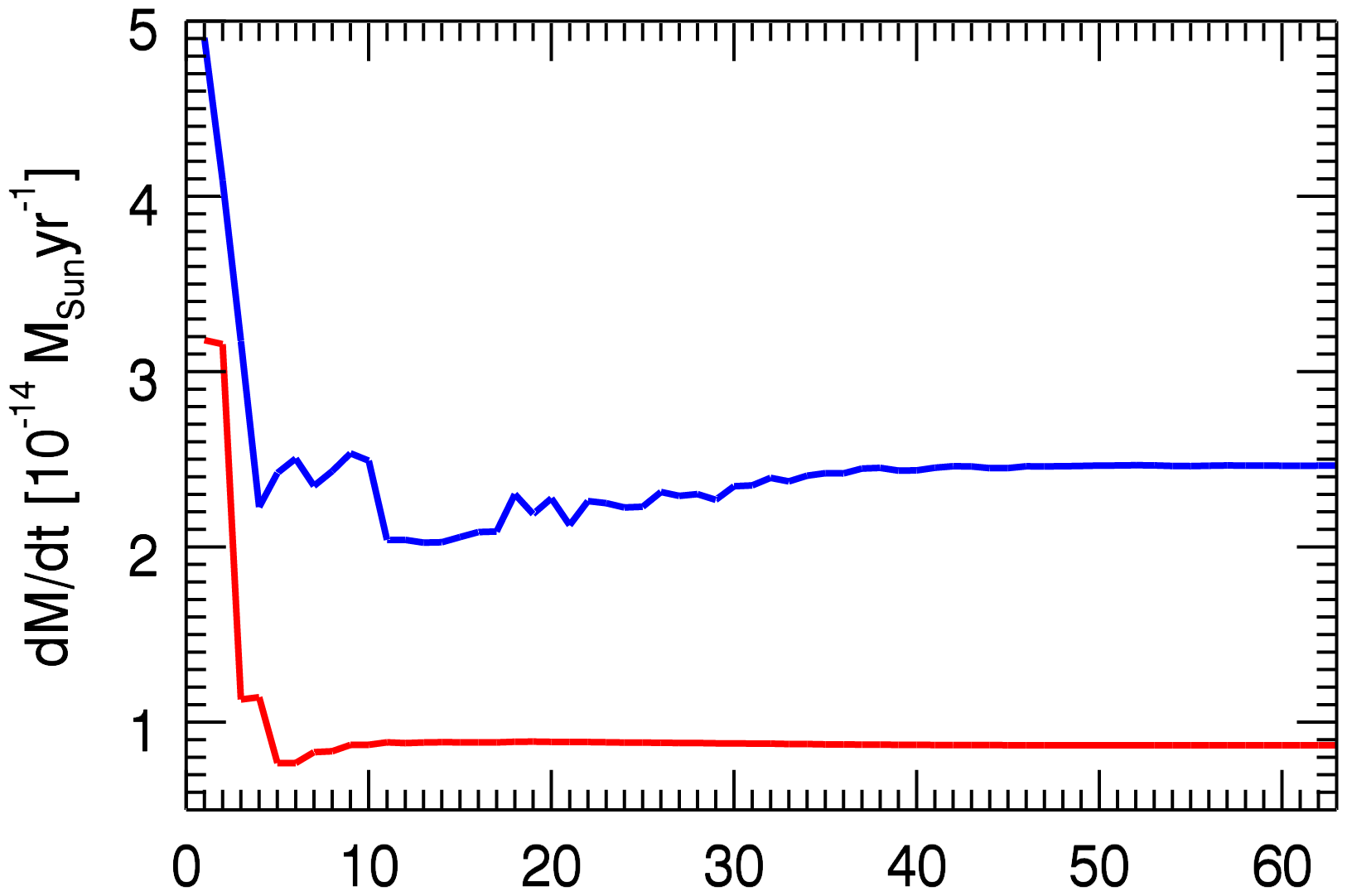}
	\includegraphics[width=70mm]{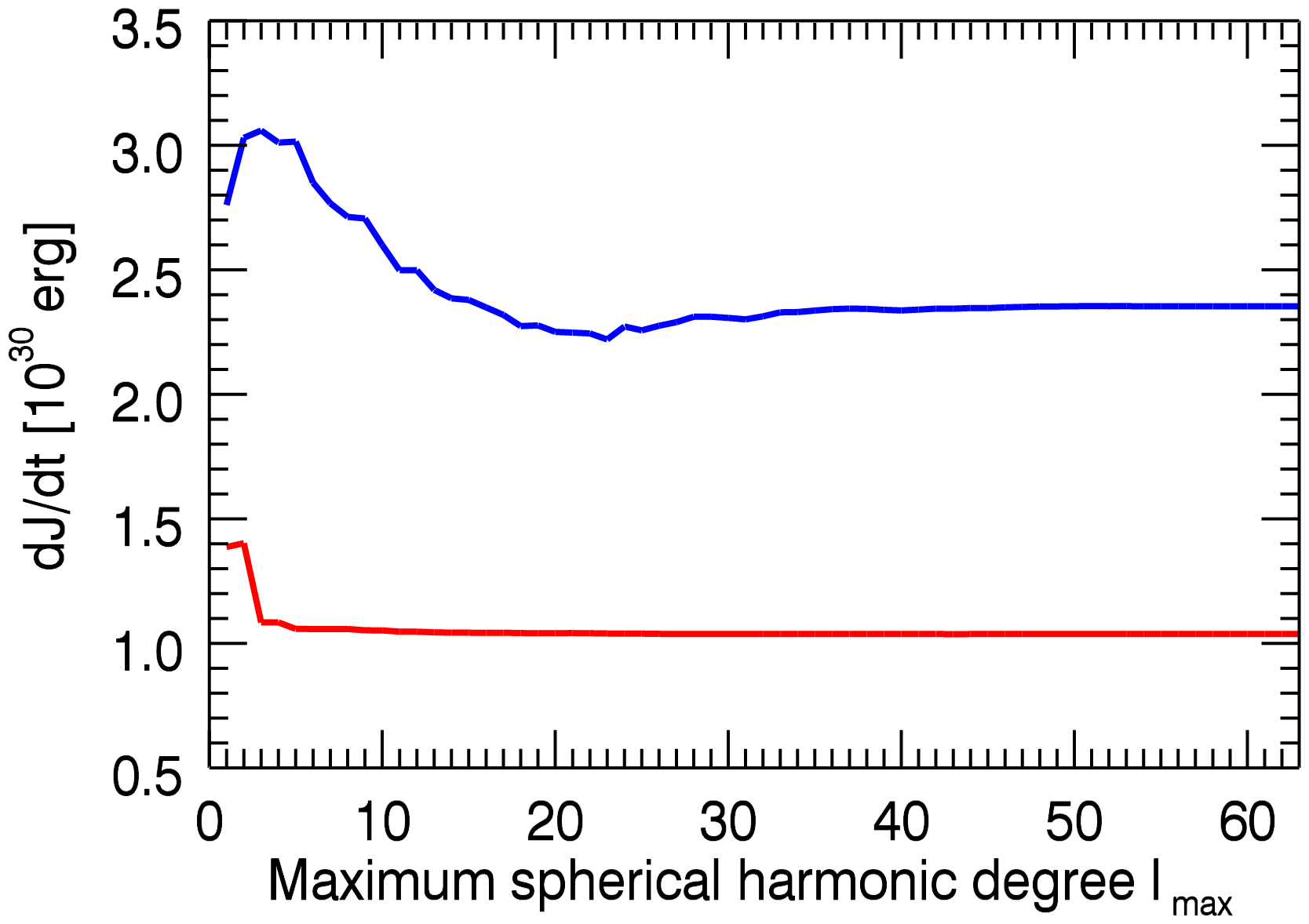}
\caption{Variation with maximum spherical harmonic degree $\ell_{\rm max}$ of the average wind speed at the Earth's orbit, the total mass loss rate and the total angular momentum loss rate.}
\label{wind}
\end{center}
\end{figure}
%

Fig. \ref{B_time} shows the surface flux and the open flux for maps sampled at different $\ell_{max}$ values. Fig. \ref{B_ell} shows, for two example magnetograms (one close to solar maximum and the other close to solar minimum), the distribution of power in the magnetic field at different lengthscales. The top panel shows that when the Sun is at its most active, the peak power is around $\ell_{max}=13$, ie the scale size of the magnetic bipoles captured by the magnetograms shown in Fig. \ref{surf_maps} (see also \citet{vidotto_sun2016}). At higher $\ell$-values, there is progressively less power. When the Sun is inactive, the peak power is in the dipole term and there is little power beyond $\ell_{max}=5$. The bottom panel shows that in the wind-dominated regime, only the lowest-order modes survive.

\subsection{Modelling the wind}
\label{sec.wind}

For each field line (labelled $i$) the velocity $u_i $ along that field line at the Earth's orbit is given by (\citet{wang1990,arge2000})
\begin{equation}
u_i [{\rm kms}^{-1}] = 267.5 + \frac{410.0}{f_i^{2/5} } 
\end{equation}
where the expansion factor $f_i$ of any field line is given by:
\begin{equation}
f_i = \frac{r_\odot^2}{r_{\rm{s}}^2} \frac{B_i(r_\odot)}{B_i(r_{\rm{s}})}.
\end{equation}
We assume that the magnetic field expands radially beyond the source surface and determine the mass loss rate from a 1D isothermal wind solution along each field line. The requirement that the wind is trans-sonic and reaches the velocity $u_i$ at Earth then determines the field line temperature. We assume that the plasma pressure at the base of the field line is given by $p_0=\kappa_w B_0^2$ where we set the free parameter $\kappa_w$ to a value that produces the variation in the solar mass loss rate through its cycle \citep{cranmer_review_2008}.  Combined with the temperature, this base pressure determines the base density. Conservation of mass and magnetic flux requires that $\rho u /B$ is constant along each flux tube, providing the mass loss rate through a spherical surface at the Earth's orbit ($S_E$)
\begin{equation}
\dot{M}=\oint_{S_E} \rho_i u_i dS_i
\end{equation}
 where $\rho_i$ is the density at the Earth's orbit and $dS_i$ is the cross-sectional area of the flux tube.  Along each field line the Alfv\'en radius is then the location where $u(r) = B(r)/\sqrt{\mu \rho(r)}$. From this, we can estimate the total angular momentum loss rate by integrating over the Alfv\'en surface ($S_A$)
\begin{equation}
\dot{J} = \oint_{S_A} \rho (\underline{u}\cdot\underline{n}) \Omega_\star \varpi^2 dS_A
\end{equation}
where $\underline{n}$ is the outward normal, $\Omega_\star$ is the stellar angular velocity and $\varpi$ is the cylindrical radius. We note that this neglects the small term due to non-axisymmetry described in \citet{mestel_book_99}. Fig. \ref{wind} shows the effect of the surface resolution on the predicted wind speed, mass loss rate and angular momentum loss rate. 

\subsection{Modelling the X-ray emission measure}
\label{EM}


\begin{figure}
\begin{center}
	\includegraphics[width=70mm]{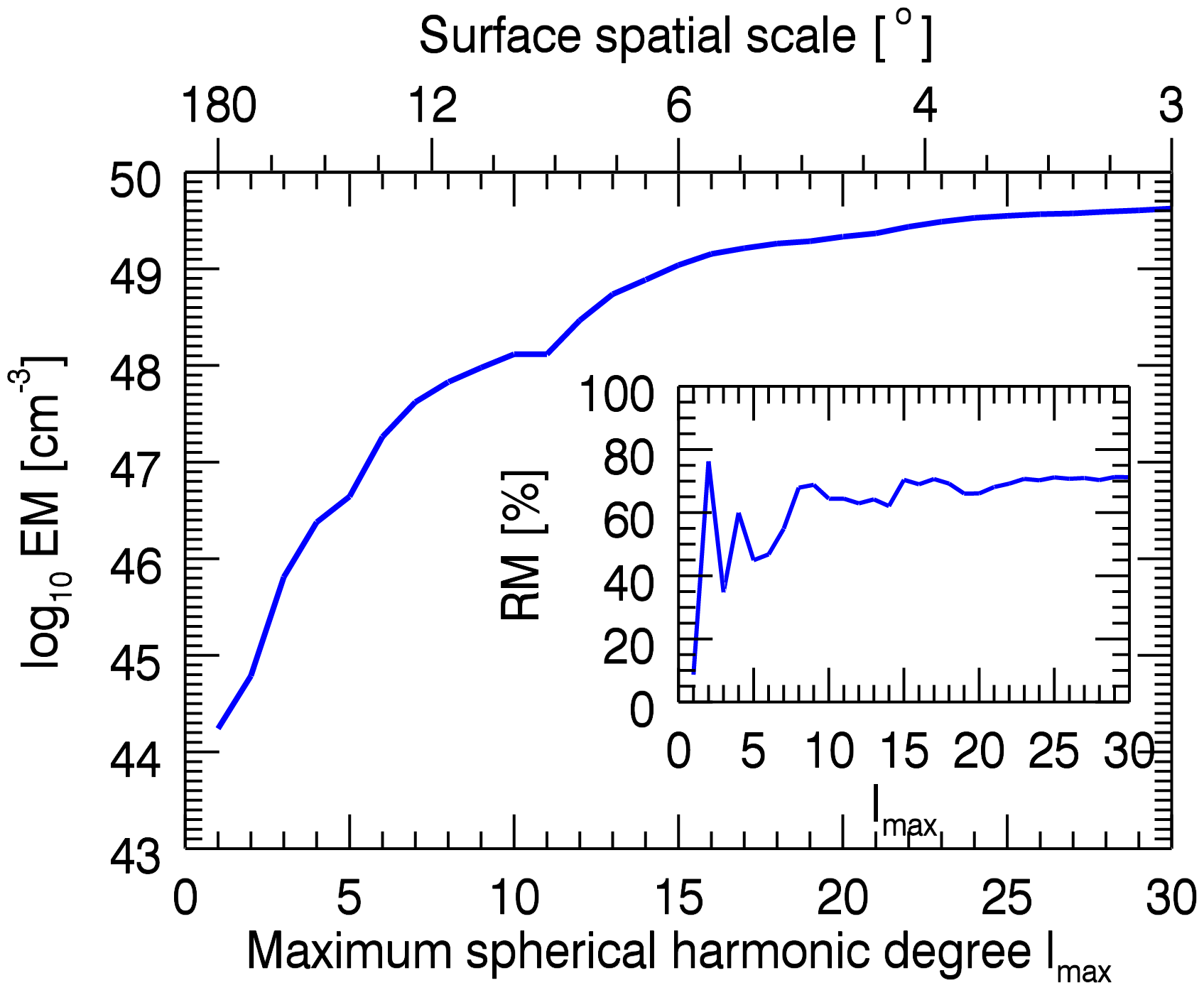}
\caption{Variation with maximum spherical harmonic degree of the emission measure for CR1851 at $10^6$K. The inset shows the rotational modulation ${\rm RM} = ({\rm EM}_{\rm max} - {\rm EM}_{\rm min})/{\rm EM}_{\rm max}$.}
\label{EM_l}

\end{center}
\end{figure}


We model the X-ray emission measure by assuming that the gas on each closed field line is in hydrostatic, isothermal equilibrium. The gas pressure is therefore $p=p_{0}e^{\frac{m}{kT}\int
g_{s}ds}$ where $g_{s} =( {\bf g.B})/|{\bf B}|$ is the component of gravity (allowing for rotation) along the field and
%
$
g(r,\theta) = \left( -GM_{\star}/r^{2} + 
                     \Omega_\star^{2}r\sin^{2} \theta,
		        \Omega_\star^{2}r\sin \theta \cos\theta 
                    \right).   
$ 
%
 The plasma pressure at the base of each field line $p_0=\kappa_c B_0^2$ where the free parameter $\kappa_c$ is fixed by the overall stellar X-ray luminosity.  In order to quantify the effect of changing the resolution of the surface magnetogram, we select the example in Fig. \ref{surf_maps}. Fig. \ref{EM_l} shows the resulting emission measures that correspond to a temperature of $10^6$ K, and their rotational modulations.

\section{Results and Discussion}
By decomposing solar magnetograms into spherical harmonics which can be truncated at different orders, we have shown the variation over the solar cycle of the various contributions to the Sun's magnetic field. As also found by \citet{derosa2012}, Fig. \ref{B_time} shows that the dipole mode has a cyclic variation that is in antiphase with the higher order modes. At each time we can also analyse the distribution of power in the magnetic field at various lengthscales (or spherical harmonic orders). At cycle maximum, this power peaks at an spherical harmonic degree determined by the spatial scale on which magnetic bipoles appear on the surface. At cycle minimum in contrast, only the lowest-order (ie largest length scale) modes contribute \citep{vidotto_sun2016}. By extrapolating this surface field out into the corona, we find that only the lowest-order modes persist out to the height at which the wind dominates over the closed corona. The flux of {\em open} magnetic field is therefore sensitive only to the lowest order modes and therefore the largest lengthscale variations of the surface field. This behaviour alone suggests that the behaviour of the stellar wind (its speed, mass loss rate and angular momentum loss rate) can be well approximated by the information in low-resolution stellar magnetograms. 

To quantify this effect, we select as an example the WSA model that predicts the 3D distribution of the solar wind speed at the Earth's orbit. We determine the variation of this wind speed with the resolution of the surface magnetogram and find that for a typical stellar surface resolution of 20$^{\rm o}$-30$^{\rm o}$, typical predicted wind speeds are within 5$\%$ of the value at full resolution. Mass loss rates and angular momentum loss rates are typically within 5-20$\%$.
%
\begin{figure}
\begin{center}
	\includegraphics[width=70mm]{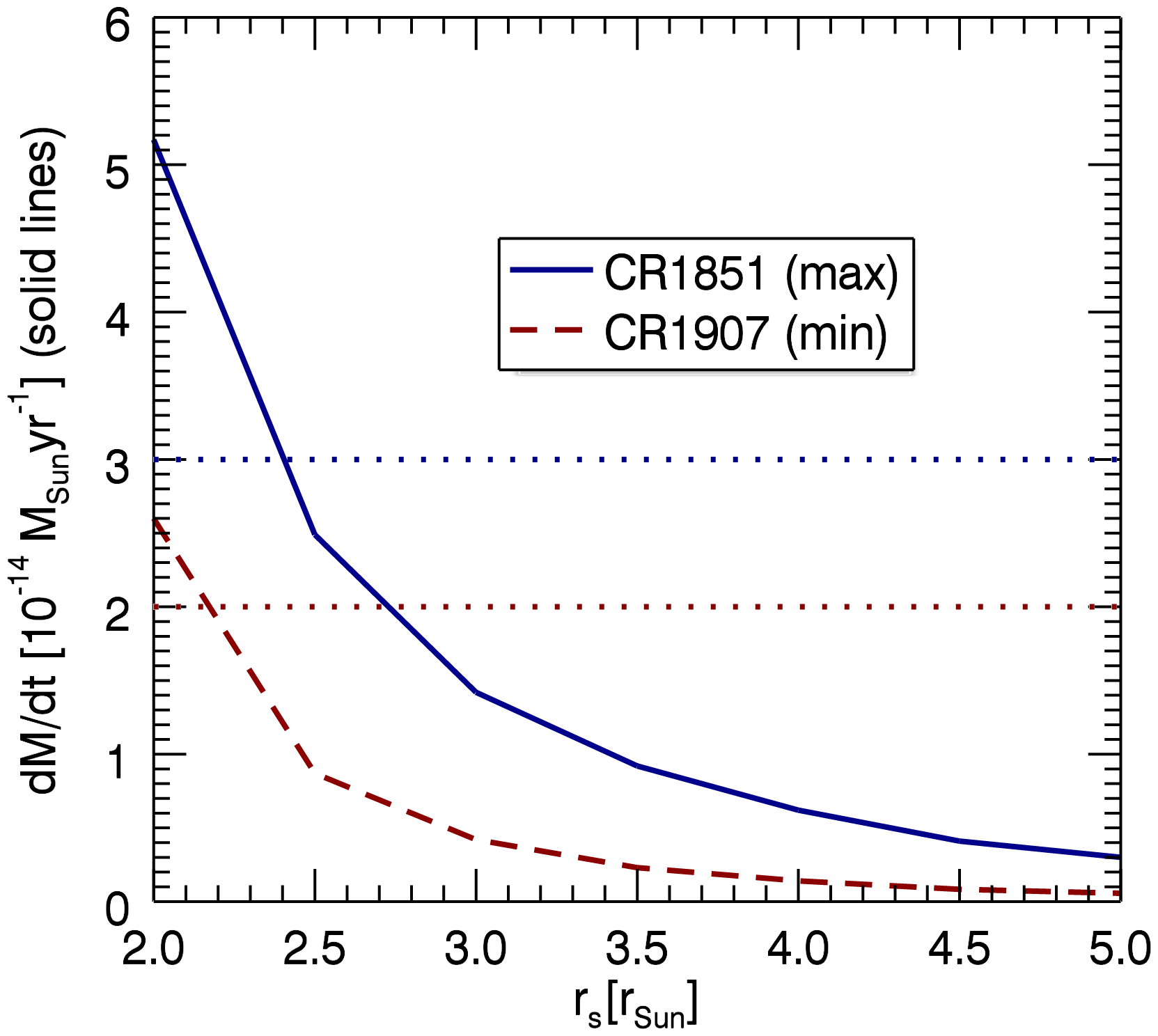}
\caption{Variation with source surface radius $r_s$ of the mass loss rate $\dot{M}$. Dotted horizontal lines show the average values at solar maximum and minimum \protect\citep{cranmer_review_2008}. An increase in $r_s$ from 2.2-2.7 $r_\odot$ between these two Carrington rotations would reproduce the observed values at these times of around $2\times 10^{-14}$M$_\odot$yr$^{-1}$. }
\label{MdotJdotRs}

\end{center}
\end{figure}

%
For comparison, we also calculate the variation of the X-ray emission measure with surface resolution. The low-resolution maps do not capture the magnetic field in the sunspots and so have lower overall field strengths and correspondingly lower emission. For a star with the same level of surface activity as the Sun, we might underestimate the emission measure by 1-2 orders of magnitude. The rotational modulation is also affected, varying from zero (the aligned dipole at solar minimum) to 80$\%$ at full resolution.

Our studies therefore suggest that wind speeds, mass, and angular momentum loss rates are insensitive to the loss of information produced by the low surface resolution of stellar magnetograms. These calculations, however, involve two free parameters  - the source surface radius $r_s$ and the constant $\kappa_w$. The mass loss rate scales linearly with $\kappa_w$. We have selected a value that reproduces the observed range of solar mass loss rates  \citep{cranmer_review_2008}. Fig. \ref{MdotJdotRs} shows the effect of varying the other free parameter $r_s$. The observed range could be reproduced with only modest adjustments of $r_s$. 

Extending this method to other stars for which surface magnetograms are available clearly requires some assumptions about the behaviour of $r_s$ and $\kappa$. Following \citet{mestel87} we suggest fixing both to the values used here for the Sun, allowing the surface magnetograms (which determine $B_0$) in addition to the stellar mass, radius and rotation rate, to govern the predicted wind properties. This produces mass loss rates for solar-like stars that compare well with those determined from fully 3D MHD wind models (See et al, 2017, submitted).


\section*{Acknowledgements}
The authors acknowledge support from STFC. 




\bibliographystyle{mnras}
\bibliography{wind_mmj} 






\bsp	
\label{lastpage}
\end{document}